\documentclass[aps,prd, 10pt, twocolumn,nofootinbib,showpacs]{revtex4-2}
\usepackage{subeqn}
\usepackage[usenames,dvipsnames]{xcolor} 

\begin{document}

\title{Spin-2 fields and cosmic censorship}
\author{Koray D\"{u}zta\c{s}}
\email{koray.duztas@okan.edu.tr}
\affiliation{Faculty of Engineering and Natural Sciences, \.{I}stanbul Okan University, 34959 Tuzla \.{I}stanbul, Turkey}
\begin{abstract}
We evaluate the possibility of the formation of naked singularities in the interaction of Kerr black holes with test massless spin-2 fields. We analyse the scattering problem for both extremal and nearly extremal black holes by incorporating the explicit form of the absorption probabilities. We show that extremal black holes become non-extremal after the interaction with the modes with both positive and negative absorption probabilities. For nearly extremal black holes the absorption probability for the challenging modes turns out to be of the order $\epsilon^5$, where $\epsilon$ parametrizes the closeness to extremality. Though the highest absorption probability for $m=2$ modes pertains to spin-2 fields, its drastically small value implies that the event horizon is preserved without the need to employ the backreaction effects. The result  can be qualitatively extrapolated to hypothetical higher integer spin fields. We also review and extend our  previous comments on Sorce-Wald method. We show that the order magnitude problems can be avoided by abandoning the non-physical parameter $\lambda$. The conditions derived by Sorce-Wald can be legitimately used to incorporate the back-reaction effects by avoiding to multiply them by $\lambda^2$, thereby preserving the magnitude of their contribution. We apply this method to the problems of over-charging R-N black holes and over-spinning Kerr black holes, in a rough analysis ignoring absorption probabilities which could apply to test bodies.
\end{abstract}
\maketitle
\section{Introduction}
A generic feature of general relativity is the development of singularities as a result of gravitational collapse  \cite{pensing}. The collapse is expected to end up in a  black hole surrounded by an  event horizon which is a  one way membrane disabling the causal connection between the singularity and the space-time outside. If the distant observers do not encounter the singularities or any effects propagating out of them, the smooth causal structure of the spacetime can be preserved at least outside the black hole region bounded by the event horizon. The weak form of the cosmic censorship conjecture asserts that singularities should always be covered by event horizons which causally disconnects them from the space-time outside \cite{ccc}. A concrete proof of the cosmic censorship conjecture appears to be elusive. 

The smooth structure of spacetimes would also be disrupted if it was possible to destroy event horizons and expose the singularity to outside observers in the interactions of black holes with test particles and fields. In the first of the thought experiments to check this possibility, Wald has shown that particles carrying sufficiently large angular momentum or charge to overspin/overcharge Kerr-Newman black holes are not absorbed by the black holes \cite{wald74}. In the following decades many similar thought experiments were constructed which involve test particles and fields \cite{hu,js,backhu,backjs,f1,gao,siahaan,magne,yuwen,higher,v1,he,wang,jamil,shay3,shay4,zeng,semiz,q1,q2,q3,q4,q5,q6,q7,overspin,emccc,natario,duztas2,mode,taubnut,kerrsen,hong,yang,bai,tjphys,khoda,ong}.
The analysis has also been extended to the asymptotically de-Sitter and anti-de Sitter cases \cite{btz,gwak3,chen,ongyao,ghosh,mtz,ext1,he2,dilat,yin,btz1,yang1,shay5,gwak4,shay6,corelli,sia2}. It turns out that cosmic censorship conjecture remains valid for perturbations satisfying the null energy condition, employing the backreaction effects if necessary. We have derived some counter-examples in alternative theories of gravity like Kerr-MOG black holes \cite{kerrmog}. However, this  points to the existence of some caveats in the underlying theory rather than the invalidity of the conjecture. 

The validity of the cosmic censorship conjecture and the laws of black hole dynamics for perturbations satisfying the null energy equation relies on the fact that there exists a lower bound for the energy of the perturbation to allow its absorption by the black hole. For test fields this lower bound corresponds to the superradiance limit where the absorption probability vanishes. One cannot derive an analogous limit for fermionic fields the energy momentum tensor of which does not satisfy the null energy condition. The absorption of the low energy modes is allowed which leads to problems regarding the validity of cosmic censorship and the laws of black hole dynamics \cite{duztas,toth,generic,spinhalf,threehalves}. The analyses for the cases that do and do not satisfy the null energy condition are fundamentally different and the results do not imply each other.  (See \cite{threehalves} for a detailed comparison of the perturbations that do and do not satisfy the null energy condition.)

In this work, we evaluate the interaction of massless spin-2 fields with Kerr black holes to search for tailored modes of the field that can overspin the black hole into a naked singularity. The energy momentum tensor for spin-2 fields satisfies the null energy condition. There exists a lower limit for the energy/frequency of a test field to allow its absorption by the black hole. The modes with a lower energy/frequency are reflected back to infinity with a larger amplitude; i.e. they exhibit superradiant scattering. Equivalently the absorption probability becomes zero and negative as the frequency decreases. Here, we incorporate the explicit form of the absorption probabilities into the analysis. We use the absorption probabilities derived by Page in his seminal work \cite{page}. In a rough analysis we would assume that the test field is entirely absorbed if the frequency $\omega$  is larger than the superradiance limit $m\Omega$. However, only a small fraction of a test field with $\omega \gtrsim m\Omega$, is absorbed by the black hole. This fraction approaches zero as the frequency approaches the superradiance limit. In particular for spin (2) fields, as $\omega$ approaches  $m\Omega$, the absorption probability decreases as $(\omega - m\Omega)^5$.  Therefore the incorporation of the absorption probabilities fundamentally changes the course of the analysis for spin (2) fields. There exist modes that could destroy the event horizon if they were entirely absorbed. This destruction would be fixed by the back-reaction effects.  However a negligible fraction of the challenging modes  is absorbed by the black hole. We show that the event horizon is preserved without invoking the back-reaction effects, when the absorption probabilities are taken into consideration.

Recently Sorce and Wald developed an alternative method and claimed to bring an ultimate solution to the problem of cosmic censorship for perturbations satisfying the null energy condition \cite{w2}. In \cite{absorp} we have scrutinized their derivation and explicitly demonstrated that it involves order of magnitude errors. The order of magnitude errors in $f(\lambda)$ defined by Sorce and Wald, are manifest when one  considers the fact that $\delta M$ is inherently a first order quantity for test particles and fields. To avoid a potential misunderstanding, we have not disputed the fact that cosmic censorship remains valid for perturbations satisfying the null energy condition. Our previous works and the current work on spin-2 fields justify this claim. In this work we extend our comments on Sorce-Wald method. We point out that the conditions derived by Sorce and Wald for the first order and second order perturbations are correct. They can be used to derive and incorporate the back-reaction effects which contribute to the interaction to second order. One can simply avoid the order of magnitude problems by abandoning $f(\lambda)$ defined by Sorce-Wald. In $f(\lambda)$,  the contribution of the back-reaction effects is multiplied by $\lambda^2$, which renders it fourth order. The correct path to follow is to derive the back-reaction effects from the Sorce-Wald condition, and substitute these contributions without multiplying them by $\lambda^2$.  Here, we use this method to re-evaluate the overcharging problem previously studied by Hubeny \cite{hu}, and the overspinning problem previously studied by Jacobson-Sotiriou for test bodies \cite{js} and D\"{u}zta\c{s}-Semiz for test fields \cite{overspin}. We adapt our previous methods to show that back-reaction effects derived from the Sorce-Wald condition can be correctly incorporated  and fix the overcharging/overspinning problems, in a rough analysis ignoring the absorption probabilities.

\section{Overspinning problem and the absorption probabilities}
In this section we describe how to incorporate the explicit form of the absorption probabilities into the overspinning problem. We start with a Kerr black hole the  mass ($M$)and the angular momentum ($J$) parameters of which satisfy:

\begin{equation}
M^2 -a^2\geq 0
\label{main}
\end{equation}
where $a=(J/M)$. We envisage that a test field is incident on the black hole from infinity. The test field is partially absorbed by the black hole and partially gets reflected back to infinity. In the test field approximation, the spacetime settles to a  new Kerr solution with modified parameters, after a sufficiently long time. If the final parameters of the spacetime violate the main inequality (\ref{main})  at the end of the interaction, the event horizon ceases to exist. The spacetime parameters represent a naked singularity which is in causal contact with the distant observers. Note that the function $(M^2-a^2)$ becomes positive, negative or zero whenever the simpler function $(M^2-J)$ is positive, negative or zero. Therefore it is customary to evaluate the simpler function $(M^2-J)$.   

The energy $(\delta M)$  and the angular momentum $(\delta J)$ of  the test field contribute to the mass and angular momentum parameters of the black hole. In a stationary and axi-symmetric spacetime, the wave-function can be separated in the form
\[ \Psi =f(r, \theta)e^{im\phi}e^{-i\omega t}
\]
The contributions to the mass and the angular momentum parameters are related by 
\begin{equation}
\delta J=\frac{m}{\omega}\delta M
\label{deltajdeltam}
\end{equation}
In \cite{absorp} we pointed out that only the fraction of the test field that is absorbed by the black hole contributes to the space-time parameters. The fact that this fraction can be very small for challenging modes, fundamentally alters the course of the analysis. The fraction that will be absorbed by the black hole is given by the absorption probability of the relevant mode. The final parameters of the space-time take the form
\begin{eqnarray}
&&M_{\rm{fin}}=M +\Gamma \delta M \nonumber \\
&&  J_{\rm{fin}}=J + \delta J= J+\frac{m}{\omega}\Gamma \delta M
\label{paramfin}
\end{eqnarray}
where $\Gamma$ is the absorption probability of the test field that is incident on the black hole, and we have expressed $\delta J$ in terms of $\delta M$ using (\ref{deltajdeltam}). The absorption probabilities for a classical scattering problem were derived by Page in a seminal work \cite{page}. For integer spin test fields, the absorption probability is given by:
\begin{eqnarray}
\Gamma&=&\left[ \frac{(l-s)!(l+s)!}{(2l)!(2l+1)!!}\right]^2 \prod_{n=1}^{l}\left[1+\left(\frac{\omega - m\Omega}{n\kappa}\right)^2 \right] \nonumber \\
&\times& 2 \left(\frac{\omega - m\Omega}{\kappa}\right)  \left(\frac{A\kappa}{2\pi}\omega \right)^{2l+1}
\label{probpage}
\end{eqnarray}
where $\omega$ is the angular frequency of the test field, $\Omega$ is the angular velocity of the event horizon, $\kappa$ is the surface gravity, and $A$ is the surface area of the black hole. Notice that the absorption probability becomes negative for modes with $\omega < m\Omega$. These modes are reflected back to infinity with a larger amplitude, in the well-known process of superradiance. The value $\omega =m\Omega $ constitutes the lower bound for the frequency of the test field for which the test field will be absorbed by the black hole. This lower bound exists for all integer spin test fields the energy momentum tensor of which satisfy the null energy condition. The fact that the low frequency modes are not absorbed by the black hole is crucial to prevent the formation of a naked singularity. The low frequency modes contribute to the angular momentum parameter much more than the mass parameter. The absorption of these modes would  lead to the violation of the main inequality (\ref{main}) and the destruction of the event horizon. For fermionic fields the absorption probability is always positive. A lower bound for the frequency to allow the absorption of the test field does not exist. This follows from the fact that the energy momentum tensor does not satisfy the null energy condition and leads to problems for the validity of cosmic censorship and the laws of black hole mechanics.

The contribution of a test field to the angular momentum parameter of the space-time is related to its azimuthal number $m$. Each field encompasses modes $l=0,1,2 \ldots$, $m=0,1,2 \ldots$. The modes with $m=0$ do not contribute to the angular momentum parameter, and the magnitude of the contribution increases as $m$ increases. Naively, one may ask the question what is the essential difference between a spin-0 and a spin-2 field regarding their contribution to angular momentum, if they can both have $m=2$? The difference lies in the absorption probabilities. For a spin-0 field the highest absorption probability pertains to $l=s=0$ modes. For the modes with $m=1$ ($l \geq 1$), the absorption probability acquires a factor $1/(3)^2=1/9$. (Check the expression in the square brackets in (\ref{probpage}).) For $m=2$ this factor becomes $1/(90)^2$. Therefore the contribution of these modes is negligible. However, for spin-2 fields the modes with $l=s=2$ have the highest absorption probability. In this case, these modes become our primary concern, and we let $m=2$ to maximize their contribution to the angular momentum parameter. Actually the highest absorption probability for $m=2$ modes pertains to spin-2 fields. In that respect, one may expect that spin-2 fields are more likely to destroy the event horizon compared to spin-0 and spin-1 case due to the absorption of $m=2$ modes. Here, we check if the absorption of $m=2$ modes of the spin-2 fields lead to overspinning by considering the explicit form of the absorption probabilities.

An explicit form for the absorption probabilities for spin (2) fields is derived by Page in \cite{page}, by substituting $l=s=2$ in (\ref{probpage}) and keeping only the lowest order terms in $\omega$. (The dominant contribution comes from $l=s$ modes)  This implies $\omega \ll m\Omega$ which renders this form of the absorption probability inappropriate for our analysis where we require that the test field is absorbed by the black hole; i.e. $\omega \geq m\Omega$. To derive a general form for the absorption probability for spin-2 fields we simply  substitute $l=s=2$ in (\ref{probpage})
\begin{eqnarray}
\Gamma&=&\left( \frac{2}{225}\right)  (\omega - m\Omega)  \left(\frac{A\omega}{2\pi} \right)^{5}   \nonumber \\
&\times& \left( \kappa^4+ \frac{5\kappa^2}{4} (\omega - m\Omega)^2 + \frac{(\omega - m\Omega)^4}{4}  \right)  
\label{probgen}
\end{eqnarray}

Using this form of the absorption probability we may proceed to evaluate the validity of cosmic censorship for extremal and nearly extremal black holes. From our previous works we know that $(\omega - m\Omega)$ must be a small number for overspinning to occur, and the surface gravity $\kappa$ is small for a nearly extremal black hole. We mentioned that the highest absorption probability for $m=2$ modes pertains to spin-2 fields. However, (\ref{probgen}) implies that the highest probability involves the fifth power of a small number which may render it negligible. 
\section{Extremal black holes}
\label{extprob}
In this section we attempt to drive an extremal black hole into a naked singularity by sending in test spin-2 fields from infinity. By definition, extremal black holes saturate the inequality (\ref{main}). 
\begin{equation}
M^2 -J=0
\label{mainext}
\end{equation}
At the end of the interaction with the test spin-2 field, the mass and the angular momentum parameters of the black hole will be modified in the form (\ref{paramfin}). We search for modes of the field  the contribution of which to the angular momentum parameter exceeds its contribution to the mass parameter so that the main inequality (\ref{paramfin}) may be violated at the end of the interaction. Since the relative contribution to the angular momentum parameter increases as the angular frequency $\omega$ decreases, it seems plausible to employ low frequency modes. However, the absorption probability (\ref{probgen}) dictates that there exists a lower bound to allow the absorption of the test field by the black hole:
\begin{equation}
\omega \geq m\Omega
\label{lowerbound}
\end{equation}
which is the well known superradiance limit. For extremal black holes (\ref{lowerbound}) takes the form
\begin{equation}
\omega \geq \frac{1}{M} \quad \left( \Omega =\frac{1}{2M} \right)
\label{lowerboundext}
\end{equation}
Note that,  we have substituted $m=2$ in  (\ref{lowerboundext})  to maximize the contribution of the test field to the angular momentum parameter. Actually, the absorption probability for the modes in the lower limit $\omega=m\Omega$, is zero. In the classical picture, the fields in these modes will be entirely reflected back to infinity without modifying the space-time parameters. The modes with $\omega< m\Omega$ get scattered back to infinity with a larger amplitude borrowing the excess energy from the angular momentum of the black hole. Therefore, we are restricted in the range $\omega> m\Omega$ if we aim to increase the angular momentum parameter.

For extremal black holes  the surface gravity, and the area are given by
\begin{equation}
\kappa=0;
 A=8\pi M^2
\end{equation}
The absorption probability (\ref{probgen})  can be simplified ($\kappa=0$)
\begin{equation}
\Gamma=\left( \frac{1}{450}\right) \left(\frac{A\omega}{2\pi} \right)^{5} (\omega - m\Omega)^5
\label{probext}
\end{equation}
The absorption probability (\ref{probext}) is drastically small for the challenging modes $\omega \gtrsim m\Omega$.  To give the reader an idea, we can consider a test spin-2 field with frequency $\omega=1.01 (1/M)$, which satisfies (\ref{lowerbound}). The absorption probability for this field is $\Gamma =2.39 \times 10^{-10}$, which implies that it is almost entirely reflected back to infinity. 

We proceed to evaluate the possibility to destroy the event horizon. After the field is partially absorbed by the black hole and partially reflected back to infinity, the final parameters of the space-time satisfy
\begin{equation}
M_{\rm{fin}}^2-J_{\rm{fin}}=\Gamma^2 (\delta M)^2 +(\delta M) \left( 2\Gamma M- \frac{2}{\omega}\Gamma \right) 
\label{deltafinext1}
\end{equation}
where we have used (\ref{paramfin}) with $m=2$, and substituted $M^2-J=0$ for the initial parameters. We may analyse the problem in 3 categories, depending on the value of $\omega$. First we evaluate the potentially challenging case $\omega > (1/M)$, for which the absorption probability is positive ($\Gamma >0$). Only the last term on the right hand side of (\ref{deltafinext1}) is negative. One observes that
\[ (\delta M) \left( 2\Gamma M- \frac{2}{\omega}\Gamma \right) >0 \left( \Gamma>0  \quad \mbox{;} \quad  \omega > \frac{1}{M} \right) \]
Therefore
\begin{equation}
M_{\rm{fin}}^2-J_{\rm{fin}}>0 \quad \mbox{for} \quad \omega > (1/M)
\end{equation}
We have started with an extremal black hole and ended up with a non-extremal black hole, with $M_{\rm{fin}}>M$ and $J_{\rm{fin}}>J$. The event horizon was preserved and the black hole has been driven away from extremality at the end of the interaction. 

Next we evaluate the case $\omega=m\Omega=(1/M)$. The absorption probability is zero and the final parameters of the space-time satisfy:
\begin{equation}
M_{\rm{fin}}^2-J_{\rm{fin}}=0 \quad \mbox{for} \quad \omega = (1/M)
\end{equation}
This test field has been entirely reflected back to infinity, without contributing to the space-time parameters. We started with an extremal black hole, and ended up with the same extremal black hole.

We can also evaluate the behaviour of the superradiant modes $\omega<m\Omega=(1/M)$. It is well-known that super-radiant modes borrow the excess energy from the rotational parameter of the black hole and drive the black hole away from extremality. Apparently these modes do not challenge the validity of cosmic censorship. We will include them in the analysis just for completeness. For these modes the absorption probability is negative ($\Gamma <0$). Again we focus on the last two terms of  (\ref{deltafinext1}) . This time the second term gives a negative contribution, while the third term gives a positive contribution.
\[ (\delta M) \left( 2\Gamma M- \frac{2}{\omega}\Gamma \right) >0 \left( \Gamma<0  \quad \mbox{;} \quad  \omega < \frac{1}{M} \right)\]
for ($\Gamma <0$) and $\omega<(1/M)$. Again, we have
\begin{equation}
M_{\rm{fin}}^2-J_{\rm{fin}}>0 \quad \mbox{for} \quad \omega < (1/M)
\end{equation}
We started with an extremal black hole and ended up with a non-extremal black hole with $M_{\rm{fin}}<M$ and $J_{\rm{fin}}<J$. The superradiant modes are scattered back with a larger amplitude. In this process both the mass and the angular momentum parameters of the black hole decrease. However, the angular momentum parameter decreases by  a larger amount. The superradiant modes borrow  most of the required energy from the rotational energy of the black hole. 

We conclude that extremal Kerr black holes cannot be overspun into naked singularities  by spin-2 fields. The event horizon is preserved and the cosmic censorship conjecture remains valid after the interaction.

\section{Nearly extremal black holes and spin-2 fields}
\label{nextprob}
We start our analysis for nearly extremal black holes by parametrizing the closeness of the black hole to extremality by a dimensionless parameter $\epsilon$. The initial parameters of the nearly extremal black hole  satisfy:
\begin{equation}
M^2-\frac{J^2}{M^2}=M^2\epsilon^2
\end{equation}
which implies
\begin{equation}
M^2-J=M^2(\epsilon^2/2)
\label{paramnext}
\end{equation}
where we assume $\epsilon \ll 1$, so that the black hole is sufficiently close to extremality to be possibly overspun by a test field. After the test field is partially absorbed by the black hole and partially reflected back to infinity, the final parameters of the space-time will take the form:
\begin{eqnarray}
\Delta_{\rm{fin}}&\equiv & M_{\rm{fin}}^2-J_{\rm{fin}} \nonumber \\
&=&M^2\frac{ \epsilon^2}{2} +\Gamma^2 (\delta M)^2 +(\delta M) \left( 2\Gamma M- \frac{2}{\omega}\Gamma \right) 
\label{deltafinnext}
\end{eqnarray}
where we have defined $\Delta_{\rm{fin}}$, and substituted $m=2$. We should search for the modes of the test field that could make $\Delta_{\rm{fin}}$ negative. (\ref{deltafinnext})  implies that  $\Delta_{\rm{fin}}$ cannot be negative  unless $\omega<(1/M)$.  We should also demand that the absorption probability for the relevant mode is positive; i.e. $\omega > m \Omega$. These are the upper and the lower bounds which constitute and interval for the frequency of the test field that could possibly overspin the nearly extremal  black hole.
\begin{equation}
m\Omega <\omega<\frac{1}{M}
\label{rangemainnext}
\end{equation}
Note that, for a nearly extremal black hole parametrized as (\ref{paramnext}),
\begin{equation}
 a^2=M^2(1-\epsilon^2)  \quad ; \quad r_{\pm}=M(1 \pm  \epsilon)
\label{arplus}
\end{equation}
$\Omega$ can be written in the form:
\begin{equation}
\Omega=\frac{a}{r_+^2+a^2}=\frac{J}{2M^2r_+}=\frac{1-\epsilon^2/2}{2M(1+\epsilon)}
\label{omeganext}
\end{equation}
Therefore the frequency of a test field that could possibly overspin the black hole is restricted in the narrow range:
\begin{equation}
\frac{1-\epsilon^2/2}{M(1+\epsilon)}<\omega <\frac{1}{M}
\label{range}
\end{equation}
where the lower bound is equal to $m\Omega=2\Omega$. In most of the previous works the explicit form of the absorption probability is not taken into consideration. It is assumed that $\Gamma \sim 1$ as long as it is positive. Here, we incorporate the explicit form of the absorption probabilities. We consider the absorption probability given in (\ref{probgen}) and evaluate it for the modes with frequency in the range (\ref{range}). To second order in $\epsilon$, the lower bound of the frequency can be written as
\[ \frac{1-\epsilon^2}{(1+\epsilon)} \sim 1- \epsilon +\epsilon^2/2
\]
where we have substituted $M=1$. (From now on we proceed with $M=1$) With these substitutions, (\ref{range}) implies that the frequency should be in the range
\begin{equation}
1- \epsilon +\epsilon^2/2< \omega <1 
\end{equation}
To first order, the maximum value of $(\omega - m\Omega)$ is equal to $\epsilon$, which was introduced to parametrize the closeness to extremality. We proceed by calculating the absorption probability for this value.
 First, note that using (\ref{arplus}), the surface gravity $\kappa$ can be written as
\[ \kappa=\frac{r_+ -r_- }{2(r_+^2 +a^2)}=\frac{\epsilon}{2(1+\epsilon)}
\]
which implies $\kappa \lesssim \epsilon$.  Let us focus on the terms appearing in paranthesis in the absorption probability (\ref{probgen})
\[  \left( \kappa^4+ \frac{5\kappa^2}{4} (\omega - m\Omega)^2 + \frac{(\omega - m\Omega)^4}{4}  \right)  \]
For ($\omega-m\Omega= \epsilon$), each term in the paranthesis is of the order $\epsilon^4$ ( Actually less than $\epsilon^4$ ). With the extra $(\omega-m\Omega)$ term outside the paranthesis, we observe that the absorption probability is less than $\epsilon^5$.
\begin{equation}
\Gamma \lesssim \epsilon^5
\end{equation}
It turns out that the absorption probability is very small for challenging modes. Let us substitute this absorption probability in (\ref{deltafinnext}) to check if it is possible to make $\Delta_{\rm{fin}}$ negative. Only the last term in (\ref{deltafinnext})  gives a negative contribution, which is
\[ \frac{2}{\omega}\delta M \Gamma\sim 2\delta M \Gamma 
\]
Substituting $\Gamma \lesssim \epsilon^5$, we observe that the absolute value of the last term can never be larger than or equal to the first term $M^2 (\epsilon^2/2)$.
\begin{equation}
M^2 (\epsilon^2/2) \gg \frac{2}{\omega}\delta M \Gamma
\end{equation}
In other words $\Delta_{\rm{fin}}$ cannot be made negative by test spin (2) fields. For nearly extremal black holes there exist frequencies that could potentially challenge the validity of the cosmic censorship conjecture. However these frequencies are restricted to a very narrow range bounded below by the limiting frequency $\omega=m\Omega$. Therefore $(\omega-m\Omega)$ is very small, which implies that the absorption probability is much smaller as it depends on $(\omega-m\Omega)^5$. A negligible fraction of these challenging modes are absorbed by the black hole which will not be sufficient to surpass the gap of width $\sim \epsilon^2$ to drive the black hole to extremality and beyond. The changes in the black hole parameters occur to sixth order considering the fact that $\delta M$ is also a first order quantity for test fields.  This follows from the involvement of the explicit form of the absorption probabilities. The cosmic censorship conjecture remains valid without employing the backreaction effects.

We can also evaluate the cases $\omega=m\Omega$ and $\omega<m\Omega$, as we did for extremal black holes. For $\omega=m\Omega$, the absorption probability is zero; $\Gamma=0$. The mass and the angular momentum parameters of the space-time remain identically the same after the interaction. For $\omega<m\Omega$, and $\Gamma <0$, the argument made in the previous section for  extremal black holes applies; i.e. $2\Gamma M-(2/\omega)\Gamma >0$. The nearly extremal black hole is driven further from extremality.
\begin{equation}
M_{\rm{fin}}^2-J_{\rm{fin}}>M^2 (\epsilon^2/2) \quad \mbox{for} \quad \omega < (1/M)
\end{equation}
It is not possible to destroy the event horizon of a Kerr black hole by test spin-2 fields. When one considers the absorption probabilities, the derivation does not require the involvement of the backreaction effects.

In the extremal case  the angular velocity of the horizon is $\Omega=(1/2M)$. In this case the superradiance limit ($\omega=m\Omega$) coincides with the upper limit of frequencies $\omega=(1/M)$, that can overspin the black hole. The range of frequencies that can be used to overspin the black hole vanishes. For nearly extremal black holes, there exists a range of frequencies ($m\Omega<\omega<(1/M)$, that could lead to overspinning. The closeness to extremality also determines the closeness of the angular velocity to the critical value $\Omega=(1/2M)$. Therefore the range of frequencies that could lead to overspinning has width $\epsilon$, which also parametrizes the closeness to extremality. The width of the range determines the maximum value of the absorption probability for the modes that would be absorbed by the black hole and lead to the destruction of the event horizon, ignoring the backreaction effects. We showed that the absorption probability for the challenging modes is of the order $\epsilon^5$, which implies that $M_{\rm{fin}}^2-J_{\rm{fin}}$ is always positive.  Had we also ignored the explicit form of the absorption probabilities and only avoided the superradiant range,  we would have derived that the test fields with frequency in the range given in (\ref{range}) could lead to the overspinning of the black hole. For $\Gamma \sim 1$, $\delta M =M \eta \sim M \epsilon$,  $M_{\rm{fin}}^2-J_{\rm{fin}}$ would vanish to first order for   $\omega=(1/M)$, and it would be negative for lower values. However, this overspinning would not be generic. We would be able to fix it by employing the backreaction effects, such as the self energy of the test field and the induced increase in the angular velocity of the black hole which were suggested by Will \cite{will}. Both of these effects would work in favour of cosmic censorship. However, by incorporating the explicit form of the absorption probabilities, we showed that the event horizon is preserved without employing backreaction effects.

\section{Sorce-Wald method}
In this section  we review our previous comments on Sorce-Wald method \cite{w2}, which first appeared in \cite{absorp}. We start with a brief review of subject. Previously, Hubeny \cite{hu} and Jacobson-Sotiriou \cite{js} attempted overcharge/overspin nearly extremal black holes by test bodies, in their well-known works. We also evaluated the same problem in a relatively well-known work involving test fields \cite{overspin}. In these thought experiments, one starts with a nearly extremal Reissner-Nordstr\"{o}m black hole parametrized in the form $M^2-Q^2 \sim M^2 \epsilon^2$ for the overcharging process, and a nearly extremal Kerr black hole parametrized as $M^2-a^2 \sim M^2 \epsilon^2$ for the overspinning process. Then one sends in a test particle or field from infinity with respective magnitudes $\delta Q \sim Q \epsilon$ and $\delta M \sim M \epsilon$. In these works the same small parameter was used to parametrize the closeness to extremality and the magnitude of the perturbation. Physically, such a choice is plausible provided that we stay in the test particle/field limit. In these works, the result is that though extremal black holes cannot, nearly extremal black holes can be overcharged or overspun neglecting backreaction effects. The overcharging and overspinning processes observed in these examples are not generic. Quantitatively one finds
\[M_{\rm{fin}}^2-Q_{\rm{fin}}^2 \sim -M_{\rm{fin}} \epsilon^2 \quad \mbox{and} \quad M_{\rm{fin}}^2-a_{\rm{fin}}^2 \sim -M_{\rm{fin}} \epsilon^2\]
which suggests that the negativeness of the right hand sides can be fixed by backreaction effects which contribute to second order in $\delta Q$ and $\delta M$. With the parametrization  $\delta Q \sim Q \epsilon$ and $\delta M \sim M \epsilon$, the backreaction effects contribute to second order in $\epsilon$. The backreaction effects involve the self energy of the test particle or field, the induced  increase in the angular velocity of the event horizon or different effects depending on the particular problem. Indeed, it was shown that backreaction effects can fix the problem of overcharging and overspinning of nearly extremal black holes by test bodies \cite{backhu,backjs}. Recently we have also shown that the induced increase in the angular velocity of the event horizon prevent the overspinning of Kerr black holes by test scalar fields \cite{kerrmog}.

Backreaction effects are usually difficult to identify and compute, and the derivations are often restricted to order of magnitude estimates. To bring an ultimate solution to this problem,  Sorce and Wald developed an alternative method to evaluate the contribution of the second order perturbations in cases satisfying the null energy condition. For that purpose Sorce and Wald first derived an expression for the minimum energy of the perturbation to allow its absorption by the black hole:
\begin{equation}
\delta M-\Omega \delta J -\Phi \delta Q \geq 0
\label{needham}
\end{equation}
Actually they have re-derived a well known relation in black hole physics. The first derivation known to this author is by Needham in 1980 \cite{needham}. The same condition was also derived by Natario, Queimeda, and Vicente in a study involving test fields in 2016 \cite{natario}. Needham's condition (\ref{needham})  determines the lower limit for the energy of perturbations satisfying the null energy condition.  It identically leads to the same results derived by Hubeny and Jacobson-Sotiriou for the lower bound of the energy in the Reissner-Nordstr\"{o}m and Kerr cases, who used different methods to derive the results. For bosonic test fields with energy $\delta M$ and angular momentum $\delta J =(m/\omega) \delta M$, it precisely gives the superradiance condition $\omega \geq m\Omega$. 

Sorce and Wald have also derived a condition for the second order perturbations.
\begin{equation}
\delta^2 M-\Omega \delta^2 J -\Phi \delta^2 Q \geq -\frac{\kappa}{8\pi}\delta^2 A
\label{condisorcewald}
\end{equation}
where $\kappa$ is the surface gravity and $A$ is the surface area of a Kerr-Newman black hole. The Sorce-Wald condition (\ref{condisorcewald}) provides a formal recipe to identify and calculate backreaction effects. However the conditions (\ref{needham}) and (\ref{condisorcewald}) should be used correctly to evaluate the interaction of black holes with test particles and fields. As we shall argue below, the incorporation of a non-physical parameter $\lambda$ leads to order of magnitude problems.

Sorce and Wald parametrize a nearly extremal black hole as usual:
\begin{equation}
M^2-Q^2-(J^2/M^2)=M^2\epsilon^2
\label{paramgen}
\end{equation}
Then, they define the function:
\begin{equation}
f(\lambda)=M(\lambda)^2-Q(\lambda)^2-J(\lambda)^2/M(\lambda)^2
\end{equation}
Next $f(\lambda)$ is expanded to second order in $\lambda$:
\begin{eqnarray}
f(\lambda)&=& (M^2-Q^2-J^2/M^2)  \nonumber \\
&+&2\lambda \left( \frac{M^4+J^2}{M^3} \delta M-\frac{J}{M^2}\delta J -Q\delta Q \right) \nonumber \\
&+&\lambda^2 \left[ \frac{M^4+J^2}{M^3} \delta^2 M-\frac{J}{M^2}\delta^2 J -Q\delta^2 Q +\frac{4J}{M^3}\delta J \delta M  \right. \nonumber \\
&-&\left. \frac{1}{M^2}(\delta J)^2+ \left( \frac{M^4-3J^2}{M^4} \right) (\delta M)^2 - (\delta Q)^2 \right]
\end{eqnarray}
Sorce and Wald proceed by imposing the Needham's condition (\ref{needham}) to ensure that the test particle or field is absorbed by the black hole. They use the Sorce-Wald condition (\ref{condisorcewald})   to eliminate the second variations.  To second order in $\lambda$, $f(\lambda)$ takes the form:
\begin{eqnarray}
f(\lambda)&\geq& M^2\epsilon^2 +\frac{2}{M^4+J^2} \left( (J^2-M^4)Q\delta Q-2JM^2\delta J \right) \lambda \epsilon \nonumber \\
&+& \frac{1}{M^2(M^4+J^2)^2}\left( (J^2-M^4)Q\delta Q-2JM^2\delta J \right)^2 \lambda^2 \nonumber \\
& &
\label{flambdafirst}
\end{eqnarray}
In attempts to overspin or overcharge black holes, the main assumption is that that the space-time metric retains its structure after the interaction, while the parameters in the metric are modified. Therefore $\delta M$ and $\delta Q$ are inherently first order quantities. Otherwise the test particle/field approximation is violated and the main assumption is not satisfied. Considering this fact, we apply an order of magnitude analysis to  $f(\lambda)$:
\begin{equation}
f(\lambda) \sim O(\epsilon^2)-O(\lambda \epsilon (\delta M) )+ O(\lambda^2 (\delta M)^2)
\end{equation}
The terms that are first order and second order in $\lambda$ contribute to $f(\lambda)$ to third and fourth order respectively.  The parameters can be expressed  in the form  $\delta M=M\eta$ and $\delta Q=Q\eta$, where $\eta \ll 1$. In general the small parameters $\epsilon$ and $\eta$ need not be equal. However, overspinning and overcharging occurs around $\eta \sim \epsilon$ in the thought experiments conducted by Hubeny, Jacobson-Sotiriou, and D\"{u}zta\c{s}-Semiz. In that case $f(\lambda)$ takes the form: 
\begin{equation}
f(\lambda)= M^2\epsilon^2-O(\lambda \epsilon^2 )+ O(\lambda^2 \epsilon^2) 
\label{problem1}
\end{equation}
The claim of Sorce and Wald is that $f(\lambda)$ becomes negative for the first order terms in $\lambda$, then it becomes positive again due to the contribution of the second order terms in $\lambda$, in accord with the previous results. However, a simple order of magnitude analysis reveals that $f(\lambda)$ is simply equal to $M^2 \epsilon^2$ to second order. Contrary to the claim of Sorce and Wald, the terms that are first order in $\lambda$ cannot make $f(\lambda)$ negative since their contribution is of the order $(\lambda \epsilon^2)$, and the terms that are second order in $\lambda$ cannot fix anything as their contribution is actually of the order $O(\lambda^2 \epsilon^2)$. Apparently  $f(\lambda)$ defined by Sorce and Wald does not reproduce the previous results by  Hubeny, Jacobson-Sotiriou, and D\"{u}zta\c{s}-Semiz. This is the order of magnitude problem we referred to in \cite{absorp}. The order of magnitude problem in $f(\lambda)$ is a brute algebraic fact, not subject to controversy.

The main contribution of Sorce and Wald in \cite{w2} is the derivation of the Sorce-Wald condition  (\ref{condisorcewald}) for the second order perturbations. However, the backreaction effects described by this condition are rendered ineffective as they are multiplied by $\lambda^2$ in $f(\lambda)$. To elucidate this point, consider the second order perturbation for a Kerr black hole perturbed by a charged particle, derived in \cite{w2}.
\begin{equation}
\delta^2 M \geq \frac{(\delta Q)^2}{2M}
\end{equation}
The result appears correct as it suggests   that the  backreactions contribute to the interaction to second order; i.e. the order $(\delta Q)^2$. However, by introducing $f(\lambda)$, one is forced to multiply the contribution of the backreaction effects by the square of an extra non-physical parameter. This renders the effect of the backreactions to become fourth order and effectively vanish. We argue that the backreaction effects can be derived from the Sorce-Wald condition (\ref{condisorcewald}), and legitimately incorporated into the analysis preserving the magnitude of their contribution. For that purpose, one should simply avoid multiplying their magnitude by $\lambda^2$.
\section{The correct application of Sorce-Wald conditions}
\label{applysw}
We pointed out that the conditions derived by Sorce and Wald are correct and it is possible to use these conditions to evaluate the possibility to destroy the event horizons in Wald type problems. For that purpose one simply abandons the non-physical parameter $\lambda$. One can define the function:
\begin{eqnarray}
\Delta_{\rm{fin}}&=&(M+\delta M+\delta^2 M)^2-\frac{(J+\delta J+\delta^2 J)^2}{(M+\delta M+\delta^2 M)^2} \nonumber \\
&-& (Q+\delta Q+\delta^2 Q)^2
\end{eqnarray}
as it is done by Semiz and D\"{u}zta\c{s}. One then imposes the Needham's condition (\ref{needham}) to ensure that the test particle or field is absorbed by the black hole. Then one can use the Sorce-Wald condition (\ref{condisorcewald}) to evaluate the impact of the backreaction effects. If $\Delta_{\rm{fin}}$ can still be made negative, one can conclude that the event horizon can be destroyed.

Let us apply this procedure to the overcharging problem studied by Hubeny. We start with a Reissner-Nordstr\"{o}m black hole which satisfies $M^2-Q^2=M^2\epsilon^2$. We send in a test particle from infinity with energy $\delta M$ and charge $\delta Q$. The lowest energy for the test particle which would allow its absorption by the black hole is given by the Needham's condition (\ref{needham}):
\[
\delta M=\frac{Q(\delta Q)}{r_+}=\frac{Q(\delta Q)}{M}(1-\epsilon + \epsilon^2)
\]
where we have substituted $r_+=M(1+\epsilon)$. At the end of the interaction $\Delta_{\rm{fin}}$ takes the form:
\begin{equation}
\Delta_{\rm{fin}}=(M+\delta M+\delta^2 M)^2- (Q+\delta Q+\delta^2 Q)^2
\end{equation}
The contribution of the backreaction effects to $\Delta_{\rm{fin}}$ is:
\begin{equation}
2(M\delta^2 M-Q\delta^2 Q)
\end{equation}
Ignoring the backreaction effects $\Delta_{\rm{fin}}$ takes the form:
\begin{equation}
\Delta_{\rm{fin}}=M^2 \epsilon^2 -2Q(\delta Q)\epsilon +O(3)
\label{deltafinQ}
\end{equation}
note that both terms in (\ref{deltafinQ}) are second order, since $\delta Q$ is a first order quantity for test particles and fields. We parametrize it as $\delta Q= Q\eta$. We observe that $\Delta_{\rm{fin}}$ becomes negative for $\eta > (\epsilon/2)$. In particular  for $\eta \sim \epsilon$ one finds:
\begin{equation}
\Delta_{\rm{fin}}=-M^2 \epsilon^2 =-Q^2 \epsilon^2
\end{equation}
which implies that the black hole is overcharged into a naked singularity if one ignores the backreaction effects. Now we incorporate the backreaction effects by using the Sorce-Wald condition  (\ref{condisorcewald}). For a Reissner-Nords\"{o}m black hole perturbed by a test object with charge $\delta Q$, one  derives that:
\begin{equation}
\delta^2 M- \frac{Q}{r_+} \delta^2 Q \geq \frac{(\delta Q)^2}{M}
\end{equation}
The derivation is given in the Appendix. Using the Sorce-Wald condition, the contribution of the backreaction effects is directly calculated:
\begin{equation}
2(M\delta^2 M-Q\delta^2 Q) \geq 2(\delta Q)^2
\end{equation}
The backreaction effects contribute to $\Delta_{\rm{fin}}$  to second order. As we argued above, this contribution should not be multiplied by the square of an extra parameter, in which case it would vanish.
Taking the effect of backreactions into the account one derives that:
\begin{equation}
\Delta_{\rm{fin}}=M^2 \epsilon^2 -2Q(\delta Q)\epsilon+2(\delta Q)^2
\end{equation}
When one incorporates the backreaction effects and substitutes $\delta Q= Q\eta$, one observes that $\Delta_{\rm{fin}}$ cannot  become negative since the expression:
\[ 2\eta^2 -2\eta \epsilon +\epsilon^2
\]
has no real roots for $\eta$. $\Delta_{\rm{fin}}$  attains its minimum value at the critical point $\eta = (\epsilon/2)$, which implies
\begin{equation}
\Delta_{\rm{fin}} \geq M^2 \frac{ \epsilon^2}{2} 
\end{equation}
The overcharging problem is fixed by the employment of the backreaction effects, which contributed to second order. The magnitude of their contribution is preserved and directly incorporated into the analysis, as we avoided to multiply them by $\lambda^2$. 

We can also apply the corrected form of Sorce-Wald method to the overspinning problem previously studied by Jacobson-Sotiriou and D\"{u}zta\c{s}-Semiz for test bodies and test fields, respectively. As usual we start with a Kerr black hole satisfying:
\[ M^2-\frac{J^2}{M^2}=M^2 \epsilon^2
\]
Again we define $\Delta_{\rm{fin}}$:
\begin{equation}
\Delta_{\rm{fin}}=(M+\delta M+\delta^2 M)^2-\frac{(J+\delta J+\delta^2 J)^2}{(M+\delta M+\delta^2 M)^2} 
\end{equation}
We want to check if $\Delta_{\rm{fin}}$ can be negative at the end of the interaction. For that purpose, it suffices to check the sign of the simpler function:
\begin{equation}
\Delta_{\rm{fin}}'=(M+\delta M+\delta^2 M)^2-(J+\delta J+\delta^2 J)
\end{equation}
We drop ``prime'', and  proceed with the modified form of  $\Delta_{\rm{fin}}$. Note that the initial parameters satisfy
\[ J^2=M^4(1-\epsilon^2) \rightarrow M^2 -J=M^2 (\epsilon^2 /2)
\]
where we used $(1-\epsilon^2)^{1/2} \simeq (1-\epsilon^2/2)$. Also note that for a nearly extremal black hole
\begin{equation}
\Omega=\frac{a}{r_+^2 +a^2}=\frac{J}{M(2Mr_+)}=\frac{(1-\epsilon^2/2)}{2M(1+\epsilon)}
\label{Omegakerr}
\end{equation} 
Next we impose the Needham's condition to maximize the contribution to the angular momentum parameter for a test particle or field that is absorbed by the black hole.
\begin{equation}
\delta J=\frac{\delta M}{\Omega}=2M \delta M(1+\epsilon+\epsilon^2 /2)
\end{equation}
We can evaluate  $\Delta_{\rm{fin}}$, for this perturbation
\begin{eqnarray}
\Delta_{\rm{fin}} &=&M^2 \frac{\epsilon^2}{2} +(\delta M)^2 -2M(\delta M) \epsilon -M\delta M \epsilon^2 \nonumber \\
 &+& (2M\delta^2 M -\delta^2 J)
\end{eqnarray}
First we ignore the contribution of the backreaction effects and evaluate the possibility to make  $\Delta_{\rm{fin}}$ negative.  We use the same parametrization as the overcharging case $\delta M=M\eta$. $\Delta_{\rm{fin}}$ takes the form:
\begin{equation}
\Delta_{\rm{fin}}=M^2 \left( \frac{\epsilon^2}{2} +\eta^2 -2\eta\epsilon+ O(3) \right)
\label{overspineta}
\end{equation}
The expression in (\ref{overspineta}) has two roots for $\eta$:
\begin{equation}
\eta_{1,2}=\epsilon \left( 1 \pm \frac{1}{\sqrt{2}} \right)
\label{etaroots}
\end{equation}
$\Delta_{\rm{fin}}$ becomes negative between these two roots for $\eta$. It attains its minimum value exactly at $\eta=\epsilon$, which equals
\begin{equation}
\Delta_{\rm{fin-min}}=\Delta_{\rm{fin}}(\eta=\epsilon)=-M^2\frac{\epsilon^2}{2}
\label{deltafinmim}
\end{equation}
Now, we incorporate backreaction effects. The Sorce-Wald condition for a Kerr black hole yields:
\begin{equation}
\delta^2 M-\Omega \delta^2 J \geq \frac{(\delta J)^2}{4M^3}
\label{sorcewaldkerr}
\end{equation}
The derivation of (\ref{sorcewaldkerr}) is also given in the Appendix. The expression on the right-hand-side of (\ref{sorcewaldkerr}) is identical with the self-energy derived from Will's argument \cite{will} based on the induced increase in the angular velocity of the event horizon. 
\[ \delta \Omega =\frac{\delta J}{4M^3} \quad ; \quad E^{1}_{\rm{self}} =\frac{(\delta J)^2}{4M^3}
\]
We previously used these backreaction effects in the overspinning problems involving bosonic and fermionic fields \cite{kerrmog,spinhalf,threehalves}. The fact that the two results coincide lends credence to the validity of the methods to derive the backreaction effects.

For a nearly extremal black hole $\Omega$ is slightly less than $1/2M$. Using this with the Sorce-Wald condition (\ref{sorcewaldcondikerr}), we evaluate the contribution of the second order perturbations; i.e. the backreaction effects.
\begin{equation}
2M \delta^2 M-\delta^2 J \geq \frac{(\delta J)^2}{2M^2}
\label{backkerr1}
\end{equation}
We would like to calculate the backreaction effects for $\eta=\epsilon$, in which case $\Delta_{\rm{fin}}$ acquires its minimum value given in (\ref{deltafinmim}). If the backreaction effects can fix the overspinning problem for the minimum value of $\Delta_{\rm{fin}}$, we can conclude that it is not possible to destroy the event horizon. Substituting $\delta M=M\eta=M\epsilon$, one derives:
\begin{eqnarray}
&& \delta J= 2M^2 \epsilon \left( 1+\epsilon + \epsilon^2 /2 \right) \nonumber \\
&& (\delta J)^2=4M^4 \epsilon^2 + O(3)
\end{eqnarray}
Substituting this value in (\ref{backkerr1})
\begin{equation}
2M \delta^2 M-\delta^2 J \geq 2M^2 \epsilon^2
\label{backkerr2}
\end{equation}
which implies that the minimum value of $\Delta_{\rm{fin}}$ becomes positive when one incorporates the backreaction effects. Namely:
\begin{equation}
\Delta_{\rm{fin}} \geq (3/2)M^2\epsilon^2
\label{deltafinmimall}
\end{equation}
The backreaction effects contributes to $\Delta_{\rm{fin}}$ to second order, and fixes the overspinning problem. Note that $\eta \sim \epsilon$ is not a simply a ubiquitous choice. Overspinning occurs in a small range around $\eta \sim \epsilon$ and $\Delta_{\rm{fin}}$ acquires its minimum value at the critical point $\eta=\epsilon$.  For this problem $f(\lambda)$ takes the form (\ref{problem1}), and equals to $M^2 \epsilon^2$ to second order. The first order and second order perturbations ( $\delta M$ and $\delta^2 M$) do not contribute to $f(\lambda)$. In that respect $f(\lambda)$ defined by Sorce and Wald does not convey any information about the interaction of the black hole with test particles and fields. Therefore one should avoid using $f(\lambda)$, and directly incorporate the backreaction effects, following the line of research developed by Semiz and D\"{u}zta\c{s}.

In our previous works, we have received comments claiming that Sorce and Wald may not have missed the order of 
magnitude errors in $f(\lambda)$, but the Sorce-Wald method applies to more general cases with $\lambda -\epsilon$ having different orders. First we should note that assigning different orders to $\lambda$ and $\epsilon$ does not change the fact that $\lambda \epsilon^2 \ll \epsilon^2$. Moreover it would be a logical mistake to identify $f(\lambda)$ as general, while it clearly does not represent the interaction in the relevant range $\eta \sim \epsilon$ where overspinning/overcharging occurs.
\section{Conclusions}
In this work we have evaluated the possibility to destroy the event horizons of Kerr black holes by sending in test spin-2 fields from infinity. Conventionally, the superradiant modes are excluded in the scattering problems involving test fields. However the explicit form of the absorption probabilities is ignored for regular modes. Namely, the absorption probability is assumed to be equal to unity, as long as it is positive. Here, we have incorporated the explicit form of the absorption probabilities into the analysis.

First, we considered the interaction of  test fields with extremal black holes. We showed that for regular modes we ended up with a nearly extremal black hole with $M_{\rm{fin}}>M$ and $J_{\rm{fin}}>J$. Both parameters increase in such a way that the black hole is driven away from extremality. We also analysed the superradiant modes. In this case the final parameters of the space-time satisfy $M_{\rm{fin}}<M$ and $J_{\rm{fin}}<J$. In this case both parameters decrease. However the angular momentum decreases by a larger amount. Again, we end up with a nearly extremal black hole. In the limiting case $\omega=m\Omega$, the absorption probability is zero. The field is entirely reflected back to infinity. The background parameters of the space-time remain identically the same after the interaction. In the conventional approach the modes with the limiting frequency are accepted as the most challenging modes. However, when one considers the fact that the absorption probability of these modes is zero, the analysis of these modes become trivial.

We analysed the case of nearly extremal black holes defining the closeness to extremality to be second order in the small parameter $\epsilon$. We derived a range of frequencies that could overspin the nearly extremal black hole if the absorption probability was of the order of unity. This is a narrow range of width $\sim \epsilon$ bounded below by $\omega=m\Omega$ and above by $\omega=(1/M)$. We mentioned that the highest absorption probability for $m=2$ modes pertains to spin-2 fields. In that respect one may expect spin-2 fields to be a better candidate for the possibility to destroy the event horizon. However, we showed that the maximum value of the absorption probability for these challenging modes is of the order $\epsilon^5$. A very small fraction of the test field is absorbed by the black hole.  Therefore the contribution to the mass and the angular momentum parameters is very small. It is not possible to surpass the gap of width $\sim \epsilon^2$ to drive the black hole to extremality or beyond. When one incorporates the absorption probabilities, the cosmic censorship conjecture remains valid without the need to employ the back-reaction effects. 

The absorption probability for $m=2$ modes for spin-2 fields is even higher than hypothetical spin-3 fields since the major contribution comes from $l=s$ modes. For the same reason the highest absorption probability for $m=3$ modes pertains to spin-3 fields. The natural question here is how we can extrapolate our results for the spin-2 case to hypothetical higher integer spin fields. In particular for spin-3 fields the absorption probability for $m=3$ modes will involve a term like $(\omega-m\Omega)^7$, therefore it will be of the order $\epsilon^7$. Though this is the highest probability for $m=3$ modes, it practically implies that $m=3$ modes will not be absorbed by the black hole. Therefore the results derived in this work imply that spin-3 and higher integer spin fields do not challenge the validity of cosmic censorship. 

We also reviewed and extended our previous comments on Sorce-Wald method. Previously we stated that the function $f(\lambda)$ defined by Sorce and Wald involves order of magnitude problems, as one is forced to multiply the contribution of the second order perturbations by the square of an extra non-physical parameter $\lambda$ \cite{absorp}. Here, we pointed out that the conditions derived by Sorce and Wald correctly describe the interaction of black holes with test particles and fields. The first condition (\ref{needham}) was first derived by Needham in 1980 \cite{needham} (and independently by different authors, see e.g. \cite{natario}). The condition on the second order perturbations (\ref{condisorcewald}) is an original contribution by Sorce and Wald. The backreactions effects can be derived by using the Sorce-Wald condition and incorporated into the analysis. To remedy the order of magnitude problems, one simply abandons $f(\lambda)$ and follows the line of research developed by Semiz and D\"{u}zta\c{s}. We applied this method to the problems of overcharging Reissner-Nordstr\"{o}m black holes and overspinning Kerr black holes. First we derived the Sorce-Wald condition for Reissner-Nordstr\"{o}m and Kerr black holes. It turns out that the condition for Kerr black holes is identical with the self-energy derived from Will's argument \cite{will}, which we have used in our previous works. We showed that overcharging and overspinning is possible when one ignores backreaction effects, and backreaction effects derived from the Sorce-Wald condition fixes the problem, in accord with previous results derived by Hubeny \cite{hu}, Jacobson-Sotiriou \cite{js}, and D\"{u}zta\c{s}-Semiz \cite{overspin}. Overspinning and overcharging occurs in the range $\delta M \sim \epsilon$, for which $f(\lambda)$ defined by Sorce and Wald acquires an invariant value of $M^2 \epsilon^2$ and conveys no information about the interaction of black holes with test particles and fields. The legitimate way to make use of the Sorce-Wald conditions is to abandon $f(\lambda)$ and directly incorporate the backreactions into the analysis.

Needham's condition gives the lower bound for the energy of the perturbation to allow its absorption by the black hole. However, one cannot directly infer that the energy of the optimal perturbations should have the value of the lower bound. In a more subtle approach, one discerns that the perturbations with the lowest energy are entirely reflected back to infinity since their absorption probability is zero. For  a test field, Needham's condition implies that the lower bound for the frequency is $\omega =m\Omega$. In the classical picture this field is entirely reflected back to infinity. Therefore the absorption probability should be taken into consideration to identify the optimal perturbations. In the examples using Needham and  Sorce-Wald conditions in section (\ref{applysw}), we ignored the absorption probabilities. In that case the derivation for the overspinning problem would be identical for test bodies, test particles, and test fields with spin-0, spin-1, and spin-2. A more accurate derivation should involve the explicit form of the absorption probabilities as executed in sections (\ref{extprob}) and (\ref{nextprob}). The problem is that the absorption probability is not well-defined for test bodies and the probabilities for test fields do not apply to test particles. (The latter is a subtle point which we elucidated in \cite{q7}.) The validity of the results ignoring the absorption probabilities is restricted to test bodies for which the absorption probability appears least relevant.

Finally we would like to point out that the results for the perturbations that do and do not satisfy the null energy condition should not be confused. For perturbations satisfying the null energy condition, there exists a lower bound for the energy to allow the absorption of the test particle or the field by the black hole. The lower bound can be derived from Needham's condition (\ref{needham}) which pre-assumes that the null energy condition is satisfied.  Fermionic fields do not satisfy the null energy condition and there exists no lower bound for the energy of the perturbation to allow its absorption by the black hole. An equivalent statement is that superradiance does not occur for fermionic fields. To be more precise the absorption probability is always positive, as justified by Page's results \cite{page}. The contribution of the lower energy modes to the angular momentum or charge parameters is much larger than their contribution to the mass parameter. The absorption of these modes by the black hole leads to a generic violation of the cosmic censorship conjecture which cannot be fixed by backreaction effects \cite{duztas,toth,generic,spinhalf,threehalves}. This does not contradict  the fact that cosmic censorship remains valid for perturbations satisfying the null energy condition, which can be considered complete with the results for spin-2 fields derived in this work.

\section*{Appendix: Sorce-Wald condition for Reissner-Nordstr\"{o}m and Kerr black holes}
In this section we evaluate the Sorce-Wald condition (\ref{condisorcewald}) for a Reissner-Nordstr\"{o}m and a Kerr black hole parametrized  as (\ref{paramgen}). The surface gravity is given by  (see Equation (116) in \cite{w2}):
\begin{equation}
\kappa=\frac{M^3}{M^4(1+\epsilon^2)+J^2}\epsilon 
\end{equation}
The second order variation in the area of the black hole is given by (see Equation (113) in \cite{w2}):
\begin{widetext}
\begin{eqnarray}
\delta^2 A&=& -\frac{8\pi}{M^8 \epsilon^3} \left \{ (\delta M)^2 \left[ J^4 + (2+\epsilon^2)J^2 M^4 -M^8(1+\epsilon)(-1 +\epsilon +2\epsilon^2) \right]  \right. \nonumber \\
&+& (\delta Q)^2 \left[ M^6Q^2 +M^8(1+\epsilon)\epsilon^2 \right] + (\delta J)^2 \left[J^2M^2 + M^6 \epsilon^2 \right ] \nonumber \\
&+& (\delta M \delta J ) \left[ -2J^3M-2JM^5(1+\epsilon^2) \right] +  (\delta J \delta Q ) (2JM^4 Q) \nonumber \\
&+& \left.  (\delta M \delta Q ) \left[ -2J^2M^3Q+2M^7Q(-1+\epsilon^2) \right] \right\}
\label{areavariation}
\end{eqnarray}
\end{widetext}

First we consider a Reissner-Nordstr\"{o}m black hole perturbed by a test particle with energy $\delta M$ and charge $\delta Q$. Note that $J=0$ and $\delta J=0$ in this case. The lowest energy to allow the absorption of the particle is given by the Needham's condition (\ref{needham}):
\begin{equation}
\delta M=\frac{Q(\delta Q)}{r_+}=\frac{Q(\delta Q)}{M}(1-\epsilon+\epsilon^2)
\end{equation}
where we have used $r_+ =M(1+\epsilon)$. Using the expression for $(\delta M)$ and the parametrization $Q^2=M^2(1-\epsilon^2)$, $Q=M(1-\epsilon^2/2)$, one derives:
\[ (\delta M)^2=\frac{Q^2(\delta Q)^2}{M^2}(1-2\epsilon +3\epsilon^2)=(\delta Q)^2(1-2\epsilon +2\epsilon^2)
\]
\[ (\delta M \delta Q)=\frac{Q(\delta Q)^2}{M}(1-\epsilon +\epsilon^2)=(\delta Q)^2(1-\epsilon +\epsilon^2/2)
\]
We evaluate the terms in curly brackets in (\ref{areavariation}) for $J=0$ and $\delta J=0$. $(\delta M)^2$, $(\delta Q)^2 $ and $(\delta M \delta Q ) $ terms contribute to $\delta^2 A$. To second order their contributions are given by:
\begin{widetext}
\[-(\delta M)^2 M^8(1+\epsilon)(-1 +\epsilon +2\epsilon^2) =(\delta M)^2 M^8(1-3\epsilon^2)=(\delta Q)^2 M^8(1-2\epsilon-\epsilon^2)
\]
\end{widetext}
\[ (\delta Q)^2( M^8 (1-\epsilon^2) + M^8 \epsilon^2)=(\delta Q)^2 M^8
\]
\[ (\delta M \delta Q ) \left[ 2M^7Q(-1+\epsilon^2) \right]=(\delta Q)^2 M^8 (-2 + 2\epsilon+ 2\epsilon^2) \]
The zeroth order and the first order terms in $\epsilon$ cancel. The total contribution of the terms in the curly brackets is $(\delta Q)^2 M^8 \epsilon^2$. For the Reissner-Nordstr\"{o}m case $\delta^2 A$ reduces to:
\begin{equation}
\delta^2 A^{\rm{RN}}=-\frac{8\pi}{M^8 \epsilon^3}\left \{ (\delta Q)^2 M^8 \epsilon^2 \right\}=-\frac{8\pi (\delta Q)^2 }{\epsilon}
\end{equation}
We can evaluate the Sorce-Wald condition for Reissner-Nordstr\"{o}m black holes:
\begin{widetext}
\[ \delta^2 M - \Phi \delta^2 Q \geq \frac{-1}{8 \pi}\left( \frac{M^3}{M^4(1+\epsilon^2)}\epsilon \right ) \left( -\frac{8\pi (\delta Q)^2 }{\epsilon} \right ) \geq \frac{(\delta Q)^2}{M(1+\epsilon^2)} \]
\end{widetext}
which directly implies
\begin{equation}
\delta^2 M - \Phi \delta^2 Q \geq \frac{(\delta Q)^2}{M}
\label{sorcewaldcondirn}
\end{equation}
Next, we evaluate the Sorce-Wald condition for Kerr black holes. In this case the Needham's condition gives:
\begin{equation}
\delta M=\Omega \delta J
\end{equation}
Recall (\ref{Omegakerr}) for the angular velocity of the event horizon
\[
\Omega=\frac{(1-\epsilon^2/2)}{2M(1+\epsilon)}=\frac{1}{2M} \left( 1-\epsilon +\epsilon^2 /2 \right)
\]
This leads to:
\[
(\delta M)^2=\frac{(\delta J)^2}{4M^2} (1-2\epsilon +2\epsilon^2) \]
\[(\delta M)(\delta J)=\frac{(\delta J)^2}{2M} \left( 1-\epsilon +\epsilon^2 /2 \right)
\]
With $\delta Q=0$, $(\delta M)^2$, $(\delta J)^2$, and $\delta M \delta J$ terms contribute to $\delta^2 A$. First we focus on the the $(\delta M)^2$ terms. We substitute the $J^2$ and $J^4$ terms in the parenthesis.
\begin{widetext}
\[
(\delta M)^2 \left[M^8(1-2\epsilon^2) + (2+\epsilon^2)M^4(1-\epsilon^2) M^4 -M^8(1+\epsilon)(-1 +\epsilon +2\epsilon^2) \right]\]
\end{widetext}
which equals (to second order)
\[
\frac{(\delta J)^2}{4M^2} (1-2\epsilon +2\epsilon^2) M^8 (4-6\epsilon^2)=(\delta J)^2 M^6(1-2\epsilon +\epsilon^2/2)
\] 
Next we evaluate the $(\delta J)^2$ terms:
\[ (\delta J)^2 (J^2 M^2 +M^6\epsilon^2)=(\delta J)^2 M^6
\]
We proceed with $(\delta M) (\delta J)$ terms. First note that:
\[
J^3=M^6 (1-\epsilon^2)(1-\epsilon^2/2)=M^6 \left( 1- \frac{3}{2} \epsilon^2 \right)
\]
The contribution of the $(\delta M) (\delta J)$ terms can be calculated as:
\begin{widetext}
\[
\frac{(\delta J)^2}{2M}\left( 1-\epsilon+ \frac{\epsilon^2}{2} \right) \left[ -2M^7 \left( 1- \frac{3}{2}\epsilon^2 \right) -2M^7 (1+\epsilon^2)(1-\epsilon^2 /2) \right]
\]
\end{widetext}
which equals (to second order)
\[(\delta J)^2 M^6 (-2 + 2\epsilon)
\]
Again, the zeroth order and the first order terms in $\epsilon$ cancel. The total contribution of the terms in the curly brackets is $(\delta J)^2 M^6 (\epsilon^2/2)$. For the Kerr case $\delta^2 A$ takes the form:
\begin{equation}
\delta^2 A^{\rm{Kerr}}=-\frac{8\pi}{M^8 \epsilon^3}\left \{ (\delta J)^2 M^6 \frac{\epsilon^2}{2} \right\}
\end{equation}
We can evaluate the Sorce-Wald condition for Kerr black holes:
\[ \delta^2 M - \Omega \delta^2 J \geq \frac{-1}{8 \pi}\left( \frac{M^3}{M^4(1+\epsilon^2)+J^2}\epsilon \right ) \left( -\frac{8\pi }{M^8 \epsilon^3} \right ) \left( (\delta J)^2 M^6 \frac{\epsilon^2}{2} \right)
\]
which exactly gives
\begin{equation}
\delta^2 M - \Omega \delta^2 J \geq \frac{(\delta J)^2}{4M^3}
\label{sorcewaldcondikerr}
\end{equation}


\begin{thebibliography}{99}
\bibitem{pensing} R. Penrose, Phys. Rev. Lett. \textbf{14}, 57 {1965}.

\bibitem{ccc} R. Penrose, Rivista del Nuovo Cim. Numero specialle \textbf{1}, 252 (1969).

\bibitem{wald74}
	R.M. Wald,  Ann. Phys. (N.Y.) \textbf{82}, 548 (1974).

\bibitem{hu} V.E. Hubeny, Phys. Rev. D \textbf{59}, 064013 (1999).

\bibitem{js}
	T. Jacobson and T.P.Sotiriou,   Phys. Rev. Lett. \textbf{103} , 141101 (2009).
	

\bibitem{backhu}  S. Isoyama, N. Sago and T. Tanaka, Phys. Rev. D
\textbf{84} 84, 124024, (2011).

\bibitem{backjs} E. Barausso, V. Cardoso and G. Khanna, Phys. Rev. Lett. \textbf{105}, 261102 (2010).

\bibitem{f1}  F. de Felice and Y. Yunqiang, Class. Quantum Grav. \textbf{18}, 1235 (2001).

\bibitem{gao} S. Gao S  and Y. Zhang, Phys. Rev. D  \textbf{87}, 044028  (2013).	 

\bibitem{siahaan} H.M. Siahaan, Phys. Rev. D   \textbf{93}, 064028 (2016).

\bibitem{magne}H.M.  Siahaan, Phys. Rev. D \textbf{96}, 024016 (2017).

\bibitem{yuwen} T.Y. Yu and W.Y. Wen,  Phys. Lett. B   \textbf{781}, 713 (2018).

\bibitem{higher} B. Wu, W. Liu, H. Tang  and R.H. Yue, Int. J. Mod. Phys. A  \textbf{21}, 1750125 (2017).

\bibitem{v1} K.S. Revelar and I. Vega, Phys. Rev. D  \textbf{96}, 064010 (2017).

\bibitem{he} Y.L. He  and J. Jiang, Phys. Rev. D  \textbf{100}, 124060 (2019).

\bibitem{wang} P. Wang, H. Wu and H. Yang, Eur. Phys. J. C \textbf{79}, 572 (2019).

\bibitem{jamil}K. D\"{u}zta\c{s} and M. Jamil, Mod. Phys. Lett. A \textbf{34}, 1950248 (2019).

\bibitem{shay3} S. Shaymatov, N. Dadhich, B. Ahmedov and M.Jamil, Eur. Phys. J. C  \textbf{80}, 481 (2020).

\bibitem{shay4} S. Shaymatov and N. Dadhich, Phys. Dark Universe  \textbf{31}, 100758 (2021).

\bibitem{zeng} D. Chen and S. Zeng, Nucl. Phys. B \textbf{957}, 115089 (2020).

\bibitem{semiz} \.{I}.  Semiz,  Gen. Relativ. Gravit \textbf{43}, 833 (2011).

\bibitem{q1}
	G.E.A. Matsas and A.R.R. da Silva, Phys. Rev. Lett.  \textbf{99},
	181301 (2007).

	
\bibitem{q2}M. Richartz and A. Saa, Phys. Rev. D \textbf{78}, 081503 (2008).
	
\bibitem{q3}S. Hod, Phys. Rev. Lett. \textbf{100}, 121101 (2008).

\bibitem{q4}G.E.A. Matsas, M. Richartz, A. Saa, A.R.R. da Silva and
	D.A.T. Vanzella,  Phys. Rev. D \textbf{79}, 101502 (2009).

\bibitem{q5} M. Richartz and A. Saa,  Phys. Rev. D \textbf{84}, 104021 (2011).
	
\bibitem{q6} S. Hod,  Phys. Lett. B \textbf{668}, 346 (2008).

\bibitem{q7} \.{I}.  Semiz   and   K. D\"{u}zta\c{s},  Phys. Rev. D
  \textbf{92}, 104021 (2015).

\bibitem{overspin} K. D\"{u}zta\c{s}  and \.{I}   Semiz, Phys. Rev. D   \textbf{88}, 064043 (2013). 

\bibitem{emccc}K. D\"{u}zta\c{s}, Gen. Relativ. Gravit. \textbf{46}, 1709 (2014).

\bibitem{natario}J. Natario, L. Queimada  and R. Vicente, Class. Quantum Grav. \textbf{33}, 175002 (2016).

\bibitem{duztas2}K. D\"{u}zta\c{s}  and \.{I}   Semiz, Gen. Relativ. Gravit.  \textbf{48}, 69 (2016).

\bibitem{mode}K. D\"{u}zta\c{s},  Phys. Rev. D \textbf{94}, 044025 (2016). 

\bibitem{taubnut}K. D\"{u}zta\c{s},  Class. Quantum Grav.  \textbf{35}, 045008 (2018).

\bibitem{kerrsen}K. D\"{u}zta\c{s},  Int. J. Mod. Phys. D \textbf{28}, 1950044 (2019).

\bibitem{hong} W. Hong, B. Mu and J. Tao, Nucl. Phys. B \textbf{949}, 114826 (2019).

\bibitem{yang} D. Chen, W. Yang and X. Zeng, Nucl. Phys. B \textbf{946}, 114722 (2019).

\bibitem{bai} T. Bai, W. Hong, B. Mu and J. Tao, Commun. Theor. Phys. \textbf{72}, 015401 (2020).

\bibitem{tjphys} K. D\"{u}zta\c{s}, Turk. J. Phys. \textbf{42}, 11 (2018).

\bibitem{khoda} H. Khodabakhshi and F. Shojai, Ann. Phys. \textbf{420}, 168271 (2020).

\bibitem{ong} Y.C. Ong, Int. J. Mod. Phys. A  \textbf{35}, 2030007 (2020).

\bibitem{btz}K. D\"{u}zta\c{s}, Phys. Rev. D   \textbf{94}, 124031 (2016).

\bibitem{gwak3} B. Gwak, J. Cosmol. Astropart. Phys. \textbf{08}, 016 (2019)

\bibitem{chen} D. Chen, Eur. Phys. J. C \textbf{79}, 353 (2019).

\bibitem{ongyao} Y.C. Ong and Y. Yao, J. High Energy Phys. \textbf{10}, 129 (2019).

\bibitem{ghosh} R. Ghosh, C. Fairoos, and S. Sarkar, Phys. Rev. D  \textbf{100}, 124019 (2019).

\bibitem{mtz}K. D\"{u}zta\c{s}, M. Jamil, S. Shaymatov and B. Ahmedov,  Class. Quantum Grav. \textbf{37}, 175005 (2020).

\bibitem{ext1} R. Ghosh, A.K. Mishra, and S. Sarkar, Phys. Rev. D  \textbf{104}, 104043 (2021).

\bibitem{he2} K.J. He, G.P. Li and X.Y. Hu,  Eur. Phys. J. C \textbf{80}, 209 (2020).

\bibitem{dilat}K. D\"{u}zta\c{s} and M. Jamil,  Int. J. Geom. Methods Mod. Phys. \textbf{17}, 2050207 (2020).

\bibitem{yin} R. Yin,  J. Liang and B. Mu,  Phys. Dark Universe \textbf{32}, 100831 (2021).

\bibitem{btz1} A.K. Ahmed, S. Shaymatov and B. Ahmedov, Phys. Dark Universe \textbf{37}, 101082 (2022).

\bibitem{yang1} S.J. Yang, Y.P. Zhang, S.W. Wei and Y.X. Liu 
J.  High Energy Phys. \textbf{2022},  66 (2022).

\bibitem{shay5} S. Shaymatov and N. Dadhich, J. Cosmol.  Astropart. Phys. \textbf{2023}, 010  (2023).

\bibitem{gwak4} B. Gwak, J. Cosmol. Astropart. Phys. \textbf{10}, 012 (2021)

\bibitem{shay6}S. Shaymatov. B. Ahmedov and M. Jamil, Eur. Phys. J. C \textbf{81}, 1131 (2021).

\bibitem{corelli} F. Corelli, M. De Amicis, T. Ikeda, P. Pani, Phys. Rev. D \textbf{107}, 044061 (2023).

\bibitem{sia2} H.M. Siahaan, P.C. Tijang, Int. J. Mod.  Phys. D \textbf{32}, 2250140  (2023)

\bibitem{kerrmog} K. D\"{u}zta\c{s}, Eur. Phys. J. C \textbf{80}, 19 (2020).

\bibitem{duztas}K. D\"{u}zta\c{s}, Class. Quantum Grav.  \textbf{32}, 075003 (2015).

\bibitem{toth}G.Z. Toth, Class. Quantum Grav. \textbf{33}, 115012 (2016).

\bibitem{generic} K. D\"{u}zta\c{s}, Eur. Phys. J. C \textbf{79}, 316 (2019).

\bibitem{spinhalf} K. D\"{u}zta\c{s}, Eur. Phys. J. C \textbf{81}, 1131 (2021).

\bibitem{threehalves} K. D\"{u}zta\c{s}, Eur. Phys. J. C \textbf{83}, 567 (2023).

\bibitem{page} D.N. Page, Phys. Rev. D \textbf{13}, 198 (1976).

\bibitem{w2} J. Sorce and R.M. Wald, Phys. Rev. D  \textbf{96}, 104014 (2017).

\bibitem{absorp} K. D\"{u}zta\c{s}, Eur. Phys. J. C \textbf{81}, 49 (2021).

\bibitem{will} C.M. Will, Astrophys. J. \textbf{191}, 521 (1974).

\bibitem{needham}T. Needham, Phys. Rev. D 22, 791 (1980).


\end{thebibliography}
\end{document}